\documentclass[twocolumn,superscriptaddress,aps,prb]{revtex4-1}
\usepackage{graphicx}
\usepackage{color}
\usepackage{amsmath}
\usepackage{amssymb}
\usepackage[colorlinks=true,
            linkcolor=blue,
            urlcolor=blue,
            citecolor=blue]{hyperref}
\definecolor{bl}{rgb}{0, 0, 1.0}

\begin{document}
\title{Role of surface termination in realizing well-isolated topological surface states within the bulk band gap in TlBiSe$_2$ and TlBiTe$_2$}

\author{Bahadur Singh}
\email{bahadursingh24@gmail.com}
\affiliation{Department of Physics, Indian Institute of Technology Kanpur, Kanpur 208016, India}
\affiliation{Centre for Advanced 2D Materials and Graphene Research Centre, National University of Singapore, Singapore 117546}
\affiliation{Department of Physics, National University of Singapore, Singapore 117542}

\author{Hsin Lin}
\email{nilnish@gmail.com}
\affiliation{Centre for Advanced 2D Materials and Graphene Research Centre, National University of Singapore, Singapore 117546}
\affiliation{Department of Physics, National University of Singapore, Singapore 117542}

\author{R. Prasad}
\affiliation{Department of Physics, Indian Institute of Technology Kanpur, Kanpur 208016, India}

\author{A. Bansil}
\affiliation{Department of Physics, Northeastern University, Boston, Massachusetts 02115, USA} 

\begin{abstract}

Electronic structures associated with the flat (polar) Se/Te- or Tl-terminated surfaces of TlBiSe$_2$ and TlBiTe$_2$ are predicted to harbor not only Dirac cone states, but also trivial dangling bond states near the Fermi energy. However, the latter, trivial states have never been observed in photoemission measurements. In order to address this discrepancy, we have carried out {\it ab-initio} calculations for various surfaces of TlBiSe$_2$ and TlBiTe$_2$. A rough nonpolar surface with an equal number of Se/Te and Tl atoms in the surface atomic layer is found to destroy the trivial dangling bond states, leaving only the Dirac cone states in the bulk energy gap. The resulting energy dispersions of the Dirac states are in good accord with the corresponding experimental dispersions in TlBiSe$_2$ as well as TlBiTe$_2$. We also show that in the case of flat, Se terminated, high-index (221) and (112) surfaces of TlBiSe$_2$, the trivial surface states shift energetically below the Dirac node and become well-separated from the Dirac cone states. 
\end{abstract}

\maketitle

\section{Introduction}
Narrow gap semiconductors with non-trivial topological invariants ($\mathbb{Z}_2$)\cite{z2_fu3DTI,z2_kaneIS,mzh_RMP1,review2,Moore10} are currently stimulating intense research activity in condensed matter physics and materials science driven by their conceptual novelties as well as potential for applications. In these novel materials, known as topological insulators (TIs),\cite{Ando13,bi2se3theory_zhang,bi2te3_expchen,bi2se3_expxia,sb2te3_hsieh,TlBiSe2_theoryyan,TlBiSe2_theoryHsin,TlBiSe2_theoryemereev,TlBiSe2_theorysingh,TlBiSe2_expkuroda,TlBiTe2_expchen,GBT124_theorysingh,tetra_Hsin,all_throughput} the bulk system is insulating, but the surfaces support metallic states with linear energy dispersion like a massless Dirac fermion. These Dirac fermions possess a helical spin-texture in \textbf{k}-space, are protected by time-reversal symmetry (TRS),\cite{qshe_theory,spintexture_Roshan} and are not allowed to backscatter in the absence of magnetic impurities or other TRS breaking perturbations. These unique properties of the Dirac states in the TIs provide an exciting playground for the realization of many topological quantum phenomena in a table-top setting, and as a basis for next generation electronic devices.\cite{spintronics_ABI,silicene2013Tsai,Magnetic_monopoles,Proximity_superconductivity,Magnetic_dopingSC,TPT_TlBiSe2_Xu,TPT_TlBiSe2_Sato,topologicalfieldtheory}

Among the various known families of three dimensional (3D) TIs, Bi$_2$Se$_3$ and Bi$_2$Te$_3$\cite{bi2se3theory_zhang,bi2te3_expchen,bi2se3_expxia,sb2te3_hsieh} have been the workhorse materials, which have been used widely for investigating topological states and their properties. They possess a relatively simple crystal structure composed of stacks of five-atomic-layer-blocks or quintuple blocks (QBs). The QBs are held together by weak van der Waals forces, providing natural cleavage planes without breaking strong bonds,\cite{dangling} and are theoretically predicted to support only non-trivial Dirac cone states without the presence of trivial dangling bond states, in agreement with the angle-resolved photoemission spectroscopy (ARPES) results.\cite{bi2te3_expchen,bi2se3_expxia,sb2te3_hsieh} Along these lines, surfaces of QBs are normally assumed to be the termination surfaces in the binary Bi-based class of TIs and other layered materials. \cite{bi2se3theory_zhang,bi2te3_expchen,bi2se3_expxia,sb2te3_hsieh,GBT124_theorysingh,tetra_Hsin}

The thallium-based ternary semiconductors TlBiSe$_2$ and TlBiTe$_2$, which are of particular interest to this study, were first predicted theoretically to be 3D-TIs, before the topological character of these materials was demonstrated experimentally.\cite{TlBiSe2_theoryHsin,TlBiSe2_theoryyan,TlBiSe2_theoryemereev,TlBiSe2_expkuroda,TlBiTe2_expchen,TPT_TlBiSe2_Xu,TPT_TlBiSe2_Sato,TlBiSe2_theorysingh} Although the crystal structure of these TIs is layered, it does not possess a weakly coupled pair of layers. Angle-resolved photoemission experiments show that TlBiSe$_2$ and TlBiTe$_2$ support a single Dirac cone surface state. \cite{TlBiSe2_expkuroda,TlBiTe2_expchen,TPT_TlBiSe2_Xu,TPT_TlBiSe2_Sato} The Dirac node in TlBiSe$_2$ lies within the bulk energy gap of 0.2$-$0.3 eV, and it is well isolated from the bulk bands. As a result, the helical spin-texture of both the upper and the lower Dirac cone in TlBiSe$_2$ is accessible and, in fact, this is the first TI in which a chirality inversion from the surface state spin-texture was observed experimentally.\cite{TPT_TlBiSe2_Xu} The velocity of the Dirac carriers in TlBiSe$_2$ is higher than Bi$_2$Se$_3$.\cite{TlBiSe2_expkuroda} Furthermore, it has been reported that bulk TlBiTe$_2$ becomes superconducting with p-doping (carrier density $6 \times10^{20}$ holes/cm$^3$).\cite{TlBiTe2_super} Reference \citenum{TlBiTe2_expchen} shows that the Fermi level at this doping lies $\sim$ 150 meV below the bulk conduction bands where six leaf-like bulk hole pockets appear in the surface Brillouin zone along with the topological surface state. These results suggest that bulk superconductivity in p-type TlBiTe$_2$ originates from the six leaf-like bulk hole pockets and thus, in this state, it may be possible for the surface Dirac cone to become superconducting due to the proximity of bulk superconductivity,\cite{TlBiSe2_theoryyan} making TlBiTe$_2$ a possible host material for topological superconductivity. 

The preceding discussion makes it clear that TlBiSe$_2$ and TlBiTe$_2$ are interesting TI materials. The cleavage plane in these crystals, however, is not obvious due to the strong ionic-covalent-type bonding among the different atomic layers. Theoretical studies predict the presence of trivial dangling bond states along with the Dirac cone states over various flat Se(Te)- as well as Tl-terminated surfaces;\cite{TlBiSe2_theoryHsin,TlBiSe2_theoryemereev,TlBiSe2_theorysingh} the trivial surface states have not been observed experimentally, even though Dirac cone states are clearly visible in the bulk energy gap. Recent scanning tunneling microscopy/spectroscopy (STM/STS) and core level-photoelectron spectroscopy (CL-PES) studies of TlBiSe$_2$ \cite{TlBiSe2_PES,TlBiSe2_PES1} indicate that the flat surface assumed in theoretical modeling may not be a realistic representation of the actual cleaved surface; TlBiSe$_2$ surface is found to exhibit a complicated morphology involving a Se-terminated surface covered with islands of Tl atoms with roughly a 50:50 coverage of Se and Tl terminations. Such a surface with an equal number of Se (anion) and Tl (cation) atoms would be nonpolar like the  (001) and (110) surfaces of recently discovered topological crystalline insulator SnTe,\cite{Snte_Hsieh} and result in the saturation of the trivial dangling bond states. Despite this experimental evidence, we are not aware of any first-principles study exploring such non-polar surfaces of TlBiSe(Te)$_2$, which might also give insight into why the predicted trivial dangling bond states are not observed experimentally.

With this motivation, we have carried out systematic {\it ab-initio} electronic structure calculations using a variety of surface terminations for both TlBiSe$_2$ and TlBiTe$_2$. When we consider an essentially nonpolar surface with an equal number of Se/Te and Tl atoms in the surface layer, as proposed in Ref. \citenum{TlBiSe2_PES}, we not only obtain Dirac cone states placed clearly within the bulk band gap in both TlBiSe$_2$ and TlBiTe$_2$, but we also find that the trivial dangling bond states now disappear. Moreover, the computed Dirac band dispersions in TlBiSe$_2$ as well as TlBiTe$_2$ for the nonpolar surfaces reproduce the corresponding experimental dispersions in remarkable detail. Finally, we explore the high-index (221) and (112) surfaces of the present compounds. The trivial states on the flat (221) and (112) Se-terminated surfaces of TlBiSe$_2$ are found to be shifted energetically to lie below the Dirac node, and here also the Dirac cone states become well isolated from the bulk as well as the trivial surface states.

The organization of this article is as follows. In Sec. \ref{computations}, we present relevant computational details. Section \ref{Result_Discuss} provides the bulk crystal and band structures and identifies the mechanism of bulk band inversion in TlBiSe$_2$ and TlBiTe$_2$. The (111) surface electronic structure for various surface terminations, and the results for (221) and (112) surface terminations are also discussed. Finally, in Sec. \ref{conclusion}, we summarize the conclusions of our study.

\section{Computational details}\label{computations}

Electronic structures were calculated within the density functional theory (DFT)\cite{kohan_dft} formalism with projector augmented wave (PAW)\cite{vasp,paw} method, using the VASP (Vienna Ab Initio Simulation Package) suite of codes.\cite{vasp} The generalized gradient approximation (GGA)\cite{pbe} was used to include exchange-correlation effects. For bulk calculations, a primitive rhombohedral unit cell of four atoms with fully relaxed structural parameters from Ref. \citenum{TlBiSe2_theorysingh} was used. In order to simulate surfaces of TlBiSe$_2$ and TlBiTe$_2$ with various terminations, we employed inversion symmetric slabs with a vacuum layer of 12 \r{A} to avoid interaction between the periodically repeated slabs. The Brillouin zone sampling was done by using $\Gamma$ centered 8$\times$8$\times$8 and 8$\times$8$\times$1 \textbf{k}-meshes for the bulk and slabs, respectively. The total energies were converged to $1.0 \times10^{-6}$ eV. Since slab construction in TlBiSe$_2$ and TlBiTe$_2$ involves breaking of strong bonds,  substantial surface relaxations can be anticipated. \cite{TlBiSe2_theoryHsin,TlBiSe2_theorysingh,TlBiSe2_theoryemereev} Accordingly, all atomic positions in the slabs were relaxed until the residual forces on each atom were less than $1.0 \times10^{-3}$ eV/\r{A}.

\section{Results and discussions}\label{Result_Discuss}
\subsection{Bulk crystal and band structure}
\begin{figure}[ht!] 
\centering
\includegraphics[width=0.48\textwidth]{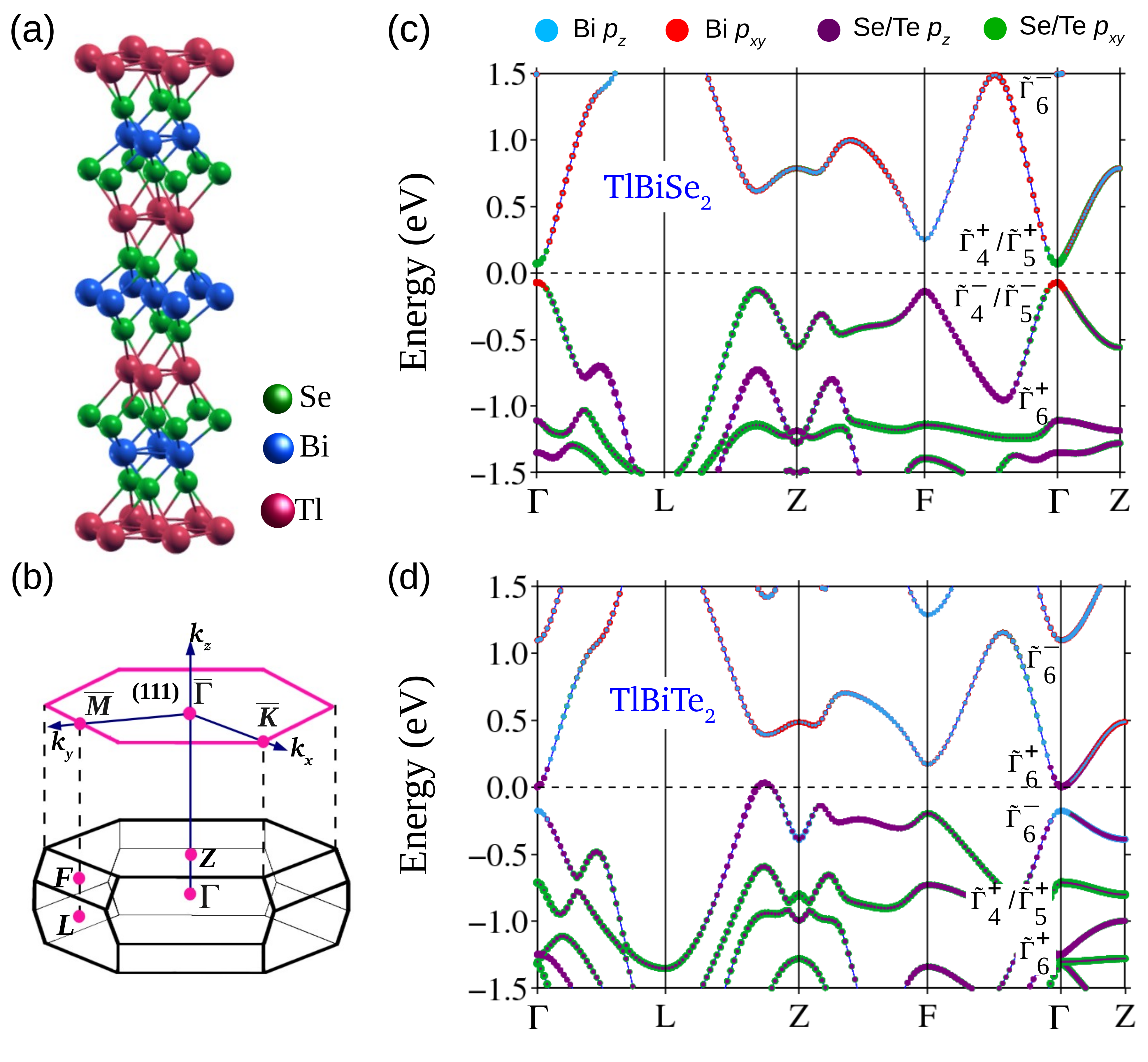} 
\caption {(a) Bulk hexagonal unit cell of TlBiSe$_2$ with 12 atomic layers. Four atomic layers with the stacking sequence -Tl-Se-Bi-Se- are repeated along the trigonal axis. (b) Bulk Brillouin zone for the primitive rhombohedral unit cell in which the four high symmetry points, $\Gamma$(0,0,0), $F$($\pi$,$\pi$,0), $L$($\pi$,0,0), and $Z$($\pi$,$\pi$,$\pi$) are marked. The projected (111) surface Brillouin zone with three high symmetry points $\overline{\Gamma}$, $\overline{M}$, and $\overline{K}$ is also shown. Bulk band structure of (c) TlBiSe$_2$ and (d) TlBiTe$_2$ along the high symmetry directions in the rhombohedral Brillouin zone. Sizes of various markers are proportional to the contribution of Bi and Se (Te) atomic states to various bands. Irreducible representations of the $D^5_{3d}$ symmetry group are indicated.\cite{TlBiSe2_theoryyan,bi2se3theory_Model1,tlbis2_Model2}}
\label{fig:bulkbs}
\end{figure}

TlBiSe$_2$ and TlBiTe$_2$ share the rhombohedral crystal structure of pseudo-Pb-chalcogenides of general type TlBiX$_2$ (X= Se, Te, or S)\cite{TlBiSe2_theoryHsin,TlBiSe2_theorysingh,TlBiSe2_theoryemereev} with four atoms per unit cell [space group $D^5_{3d}~(R\overline{3}m)$]. The conventional hexagonal unit cell of TlBiSe$_2$ is shown in Fig. \ref{fig:bulkbs}(a) as an example. It is composed of three formula units of TlBiSe$_2$ with strongly bonded, alternating cation (Bi$^+$ or Tl$^+$) and anion (Se$^-$) layers [-Tl-Se-Bi-Se-]$_n$ along the $z$ axis of the hexagonal unit cell [(111) axis of the rhombohedral unit cell]. Unlike Bi$_2$Se$_3$, where the strongly bonded QBs [-Se-Bi-Se-Bi-Se-]$_n$ are held together by weak van der Waals forces, in TlBiSe$_2$, each Tl(Bi) layer is strongly coupled with the two neighboring Se layers.\cite{TlBiSe2_theorysingh,TlBiSe2_theoryemereev} The distance between Tl and Se layers is larger ($d=3.30$ \r{A}) compared to that between Bi and Se layers ($d=2.99$ \r{A}), reflecting the stronger coupling between the latter layers. As a result, it will be natural for the crystal to cleave between the Tl and Se layers, a point to which we return below.

Bulk band structures of TlBiSe$_2$ and TlBiTe$_2$ along the high symmetry directions in the bulk Brillouin zone [see Fig. \ref{fig:bulkbs}(b)] are shown in Figs. \ref{fig:bulkbs}(c) and \ref{fig:bulkbs}(d), respectively. Both compounds exhibit a spin orbit coupling (SOC) induced band inversion at the $\Gamma$ point where the Bi $p$-states lie below the Se/Te $p$-states. This band ordering is opposite to that away from the zone center where Se/Te $p$-states are occupied and Bi $p$-states are unoccupied, resulting in non-trivial $\mathbb{Z}_2$ invariants: $\nu_0;(\nu_1\nu_2\nu_3$)=1;(000).\cite{z2_kaneIS}

In order to construct an effective low-energy Hamiltonian, it is important to identify the nature of the orbitals that control the electronic structure of a material in the vicinity of $E_F$. With this motivation, we now turn to discuss characters of the relevant states in the present compounds. The band inversion in TlBiSe$_2$ and TlBiTe$_2$ occurs around the $\Gamma$-point and therefore, we characterize these states via the irreducible representations of the $D^5_{3d}~(R\overline{3}m)$ group, \cite{TlBiSe2_theoryyan,bi2se3theory_Model1,tlbis2_Model2} which can be determined by expressing the crystal wave functions as a superposition of the appropriate atomic orbitals.\cite{bi2se3theory_Model1} It is thus straightforward to show that at the $\Gamma$-point, the Bi-$p_z$-type wave functions transform like $\widetilde{\Gamma}^{-}_{6}$(j=1/2), while the Bi-$p_{xy}$-type wave functions involve both the $\widetilde{\Gamma}^{-}_{4}$ and $\widetilde{\Gamma}^{-}_{5}$(j=3/2) representations. Similarly, the Se/Te-$p_z$- and $p_{xy}$-like states transform as $\widetilde{\Gamma}^{+}_{6}$(j=1/2) and $\widetilde{\Gamma}^{+}_{4}/\widetilde{\Gamma}^{+}_{5}$(j=3/2), respectively. Since $p_{xy}$-like states involve a combination of $\widetilde{\Gamma}^{-(+)}_{4}$ and $\widetilde{\Gamma}^{-(+)}_{5}$ representations, we have labeled these states as $\widetilde{\Gamma}^{-(+)}_{4}/\widetilde{\Gamma}^{-(+)}_{5}$ in Figs. \ref{fig:bulkbs}(c) and \ref{fig:bulkbs}(d).

The band structure of TlBiSe$_2$ in Fig. \ref{fig:bulkbs}(c) shows that the Se-$p_z$-like $\widetilde{\Gamma}^{+}_{6}$ valence levels lie below the Se-$p_{xy}$-like $\widetilde{\Gamma}^{+}_{4}/\widetilde{\Gamma}^{+}_{5}$ states, while the Bi-$p_z$-like $\widetilde{\Gamma}^{-}_{6}$ conduction levels lie above the Bi-$p_{xy}$-like $\widetilde{\Gamma}^{-}_{4}/\widetilde{\Gamma}^{-}_{5}$ levels. The band ordering in the crystal-field split valence states is the normal order with $p_z$ lying below the $p_{xy}$-like states. This order is, however, inverted in the conduction bands where $p_z$ lies above $p_{xy}$, controlling the band inversion in TlBiSe$_2$. The order of crystal field splittings of the valence as well as the conduction bands in TlBiTe$_2$ is opposite to that in TlBiSe$_2$, so that the band inversion now occurs between $\widetilde{\Gamma}^{-}_{6}$ (j=1/2) and $\widetilde{\Gamma}^{+}_{6}$ (j=1/2), see Fig. \ref{fig:bulkbs}(d). These differences reflect those in the ionicities of the underlying atomic bonds and the resulting structural distortions along the high symmetry directions in the two compounds.\cite{BiTeI_ModelSplitt}

\subsection{Surface termination and surface states}
\begin{figure}[ht!]
\centering
\includegraphics[width=0.48\textwidth]{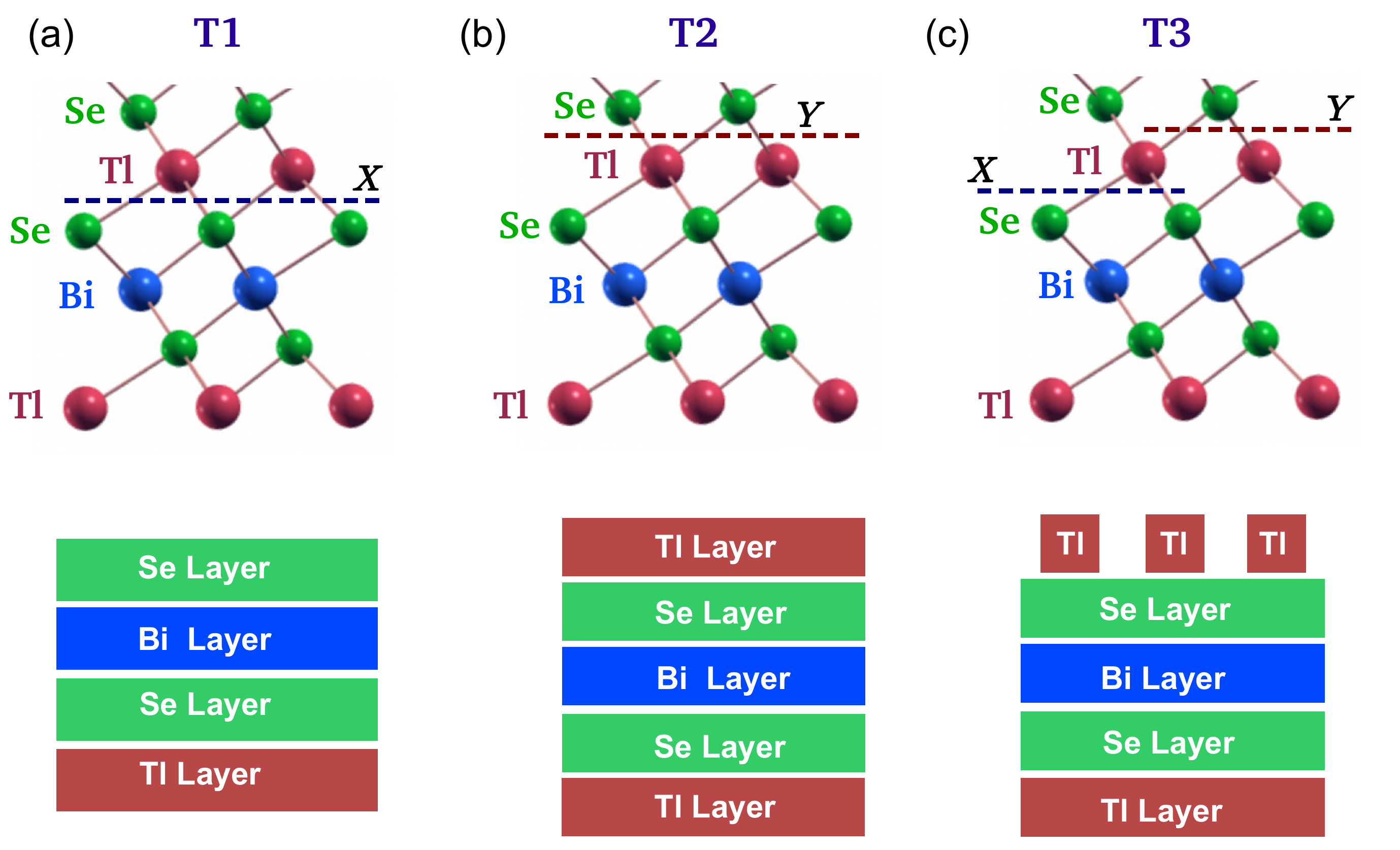} 
\caption{Three different models of surface termination in TlBiSe$_2$ are shown (top row) along with a schematic of the associated layer structures (bottom row). Dashed lines marked with {\textbf {\textit X}} and {\textbf {\textit Y}} identify two different ways of breaking the Tl-Se bonds.}
\label{fig:termination}
\end{figure}

Keeping in mind that the bonding between Tl and Se (Te) layers in TlBiSe$_2$  (TlBiTe$_2$) is weak compared to other layers, we consider three different possibilities for the (111) surface termination. These are illustrated in Fig. \ref{fig:termination} with the example of TlBiSe$_2$: (i) T1:  Tl-Se bonds are broken as shown by dashed {\textbf {\textit X}} line in Fig. \ref{fig:termination}(a). Se atoms terminate the surface with Bi atoms in the second layer. This is the most commonly employed surface termination. \cite{TlBiSe2_theoryHsin,TlBiSe2_theorysingh,TlBiSe2_theoryemereev} (ii) T2: Tl-Se bonds are broken as shown by the dashed {\textbf {\textit Y}} line in Fig. \ref{fig:termination}(b), with the Tl atoms terminating the surface. (iii) T3: Tl-Se bonds are broken such that half of the surface is like T1 while the other half is like T2 as shown in Fig. \ref{fig:termination}(c), consistent with experimental observations.\cite{TlBiSe2_PES,TlBiSe2_PES1} This surface is nonpolar with an equal number of Tl$^+$ and Se$^-$ atoms, and since the bulk stoichiometry is unbroken \textit{i.e.} $N_{\text{Tl}}=N_\text{Bi}=2\times N_\text{Se}$, similar to Bi$_2$Se$_3$ and SnTe, charge neutrality is maintained throughout the system in this case. 

Breaking of Tl-Se/Te bonds leaves unpaired electrons or unsaturated bonds, and as a result, the surface atoms in TlBiSe$_2$ and TlBiTe$_2$ will likely undergo rearrangements and/or reconstructions in achieving a stable configuration.\cite{Reconst_theory,Reconst_exp} We have examined this possibility by allowing all atomic positions to relax in our computations. However, the surface atoms for T1 as well as T2 termination in both compounds maintain their in-plane positions without a tendency for reconstructions, reflecting presumably the highly directional nature of the bonding of the surface atoms with the underlying layers. \cite{TlBiSe2_theoryHsin,TlBiSe2_theorysingh,TlBiSe2_theoryemereev} On the other hand, there are significant relaxation effects with atoms experiencing out-of-the-plane displacements; the distance between the first two top layers contracts, while that between the second and third layers expands with this pattern of alternating contraction and expansion decaying as one goes deeper into the bulk. Specifically, the distance between the first two layers in TlBiSe$_2$ (TlBiTe$_2$) contracts by 0.09 (0.11) \r{A} and 0.22 (0.19) \r{A} for T1 and T2 terminations, respectively; the corresponding increases in the distance between the second and third layers is 0.11 (0.18) \r{A} and 0.12 (0.13) \r{A} for T1 and T2. 

Turning to T3 termination, we modeled this complicated surface by taking a $2\times 2 \times 1$ supercell of a Tl-terminated slab in which we removed Tl atoms from half of the surface layer as shown in Fig. \ref{fig:termination}(c).\cite{footnote1} The resulting surface has 50:50 Se/Tl coverage, and it does not possess in-plane bulk periodicity. This is in sharp contrast to the surfaces with T1 and T2 terminations where the in-plane bulk periodicity over the surface layers is maintained, see Fig. \ref{fig:termination}. The interlayer distances in this case also show out-of-the-plane relaxations, but the size of the deviations from bulk values in TlBiSe$_2$/TlBiTe$_2$ is smaller than for T1 or T2 termination, and the interlayer spacing rapidly converges to the bulk value.
 
\begin{figure}[ht!]
\centering
\includegraphics[width=0.48\textwidth]{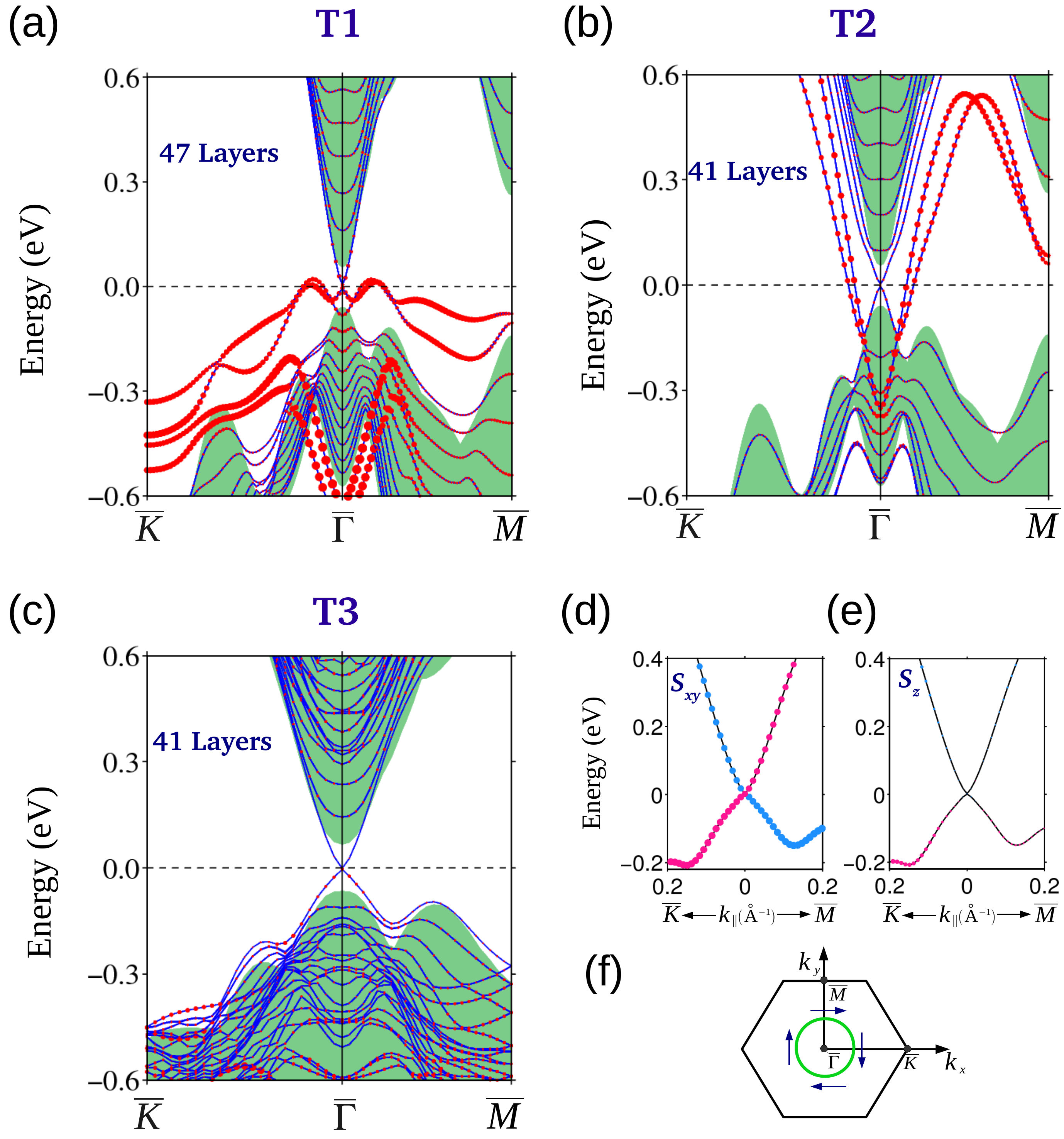} 
\caption{Band structure of the (111) surface of a symmetric slab of TlBiSe$_2$ for various surface terminations: (a) T1 termination with Se-exposed surface; (b) T2 termination with Tl-exposed surface; and (c) T3 termination with Tl- and Se-exposed surfaces (see Fig. \ref{fig:termination} for the meaning of T1-T3). Sizes of red circles correspond to the weights of the associated states in the top two atomic layers in the slabs; the background green color represents the projected bulk bands. (d) In-plane ($S_{xy}$), and (e) out-of-the-plane ($S_z$) spin-polarization of the topological states in T3 termination. Sizes of the pink and blue circles are proportional to the positive and negative values of the spin, respectively. (f) Schematic of the in-plane spin-texture (left handed) for the surface state electrons in TlBiSe$_2$.}
\label{fig:tlbise2}
\end{figure}

Band structures of fully relaxed (symmetric) slabs of TlBiSe$_2$ with various surface terminations (T1-T3) are shown in Figs. \ref{fig:tlbise2}(a)-\ref{fig:tlbise2}(c). All three terminations support Dirac cone surface states, which lie within the bulk energy gap. T1 and T2 also yield trivial surface states, which are localized in the top two atomic layers of the slab as seen from sizes of the red circles in Figs. \ref{fig:tlbise2}(a) and \ref{fig:tlbise2}(b); these are dangling bond states that arise from unsaturated bonds of the surface atoms.\cite{Reconst_exp} The T3 termination, on the other hand, does not support such trivial surface states around the Fermi level, see Fig. \ref{fig:tlbise2}(c). The absence of trivial states in T3 may be attributed to the stoichiometric slab configuration ($N_{\text{Tl}}=N_\text{Bi}=2\times N_\text{Se}$) where we have a nonpolar surface geometry with an equal number of Tl$^+$ and Se$^-$ layers. Since the dangling bond states are localized on Tl and Se surface atomic layers, the strong ionic nature of bonding between Tl and Se atoms maintains the charge balance over the surface and eradicates these states. Notably, for T1 and T2 terminations, the lower Dirac cone is quite distorted, and the Dirac node overlaps with trivial surface states. In sharp contrast, for the T3 termination, both the upper and lower Dirac cones exhibit nearly linear energy dispersion, the Dirac node lies in the middle of the bulk gap, and it is well isolated from other (trivial) surface states.

The (111) surface of TlBiSe$_2$ has been extensively studied via angle-resolved photoemission spectroscopy (ARPES), which clearly shows the presence of a Dirac-type cone,\cite{TlBiSe2_expkuroda,TlBiTe2_expchen,TPT_TlBiSe2_Xu,TPT_TlBiSe2_Sato} with its node lying 0.3-0.4 eV below the Fermi level within a bulk gap of 0.2-0.3 eV depending on details of the sample used.\cite{TlBiSe2_expkuroda,TlBiTe2_expchen,TPT_TlBiSe2_Xu,TPT_TlBiSe2_Sato} Dispersion of the Dirac cone is observed to be isotropic up to $\sim$ 200 meV above or below the node, beyond which a bulk valence band can be seen at \textbf{k} $\sim$ 0.2 $\r{A}^{-1}$ along the $\overline{\Gamma}-\overline{M}$ directions.\cite{TlBiSe2_expkuroda,TlBiTe2_expchen} No evidence of other surface states in the vicinity of the Fermi energy is found.

The preceding experimental results on TlBiSe$_2$ are in good accord with our first-principles computations based on the T3 model termination as follows. The computed Dirac cone resides within the bulk energy gap of 160 meV with the Dirac node lying roughly in the middle of this gap at the $\overline{\Gamma}$ point.  The bulk band features observed in the experiments along $\overline{\Gamma}-\overline{M}$ and $\overline{\Gamma}-\overline{K}$ directions can be seen clearly in the {\it ab-initio} energy dispersions; an example is the valence band maximum at \textbf{k} $\sim$ 0.2 $\r{A}^{-1}$ along the $\overline{\Gamma}-\overline{M}$ directions at an energy of 120 meV in Fig. \ref{fig:tlbise2}(c), which is also seen in the experiments.\cite{TlBiSe2_expkuroda,TlBiTe2_expchen} Finally, the computed electronic structure does not show the presence of any other (trivial) surface states in the bulk gap overlapping with the Dirac cone in sharp contrast to the T1 and T2 terminations, resolving a puzzling discrepancy between the earlier computations and experiments.

In order to check the stability of the T3 terminated surface, we have computed the total energies of slabs with various terminations. Note that slabs with different terminations contain different numbers of atoms and, therefore, their energies cannot be compared directly. However, the average energy of the slabs with T1 and T2 terminations can be compared with that of the T3-terminated slab as follows. If E1 is the total energy of an $N$ layer T1 slab and E2 the energy of the $N+\{1\times 2\}$ layer T2 slab, then $E^{avg}=\frac{1}{2}(E1+E2)$ is the energy of a slab with $N+\{\frac{1}{2}\times 2\}$ layers. (The multiplication factor of $2$ in the curly brackets here accounts for the inversion symmetry of the slab.) $E^{avg}$ can now be compared directly with the energy of the corresponding T3 slab with $N+\{\frac{1}{2}\times 2\}$ layers.

\begin{table}[h!] 
\centering
\caption{Total energies per atom of TlBiSe$_2$ and TlBiTe$_2$ slabs with various surface terminations. E1: 39 layers, T1-termination; E2: 41 layers, T2 termination; E3: 40 layers, T3 termination; E$^{avg} = \frac{1}{2}(E1+E2)$. Energies are given in units of eV/atom.}
\begin{tabular}{l c c c c } \hline \hline
            &  E1      &  E2     &     E3     &  E$^{avg}$ \\ 
            \hline 
TlBiSe$_2$  &   -3.871 & -3.826  &   -3.861   &  -3.848    \\
TlBiTe$_2$  &   -3.611 & -3.569  &   -3.599   &  -3.589    \\ 
\hline \hline
\end{tabular}
\label{tab:Energy}
\end{table}

Table \ref{tab:Energy} presents total energies per atom computed for slabs with various terminations. The energy per atom for the T3-termination, E3, is seen to be smaller than the corresponding average energy for T1 and T2 terminations, E$^{avg}$, indicating that T3 termination is energetically more favorable. This is consistent with the results of recent STM/STS studies on the surface morphology of TlBiSe$_2$\cite{TlBiSe2_PES,TlBiSe2_PES1} as well as the ARPES results on TlBiSe$_2$  discussed above. We thus conclude that T3 termination is a good model for the naturally occurring stable (111) surface of TlBiSe$_2$.\cite{footnote2}

A salient feature of the Dirac states is their helical spin-texture. In this connection, we have computed the spin-textures of the Dirac states by evaluating the expectation values of spin operator for three spin directions as a function of \textbf{k}.\cite{,bi2se3theory_Model1,vasp} Figures \ref{fig:tlbise2}(d) and \ref{fig:tlbise2}(e) present results for the in-plane ($S_{xy}$) and out-of-the-plane ($S_z$) spin components for the T3 termination. It can be seen from the sizes of the pink and blue circles that the Dirac states have a large in-plane spin polarization up to a momentum \textbf{k} $\sim$ 0.15 $\r{A}^{-1}$ around the Dirac-point, beyond which a finite out-of-plane spin component develops due to coupling with the bulk states. The in-plane spin-polarization has a left-handed (clockwise) chirality for the upper Dirac cone [see Fig. \ref{fig:tlbise2}(f)], but a right-handed (counterclockwise) chirality for the lower Dirac cone. These theoretical results are also in good accord with the corresponding spin-resolved ARPES experiments.\cite{TPT_TlBiSe2_Xu}  

\begin{figure}[ht!]
\centering
\includegraphics[width=0.48\textwidth]{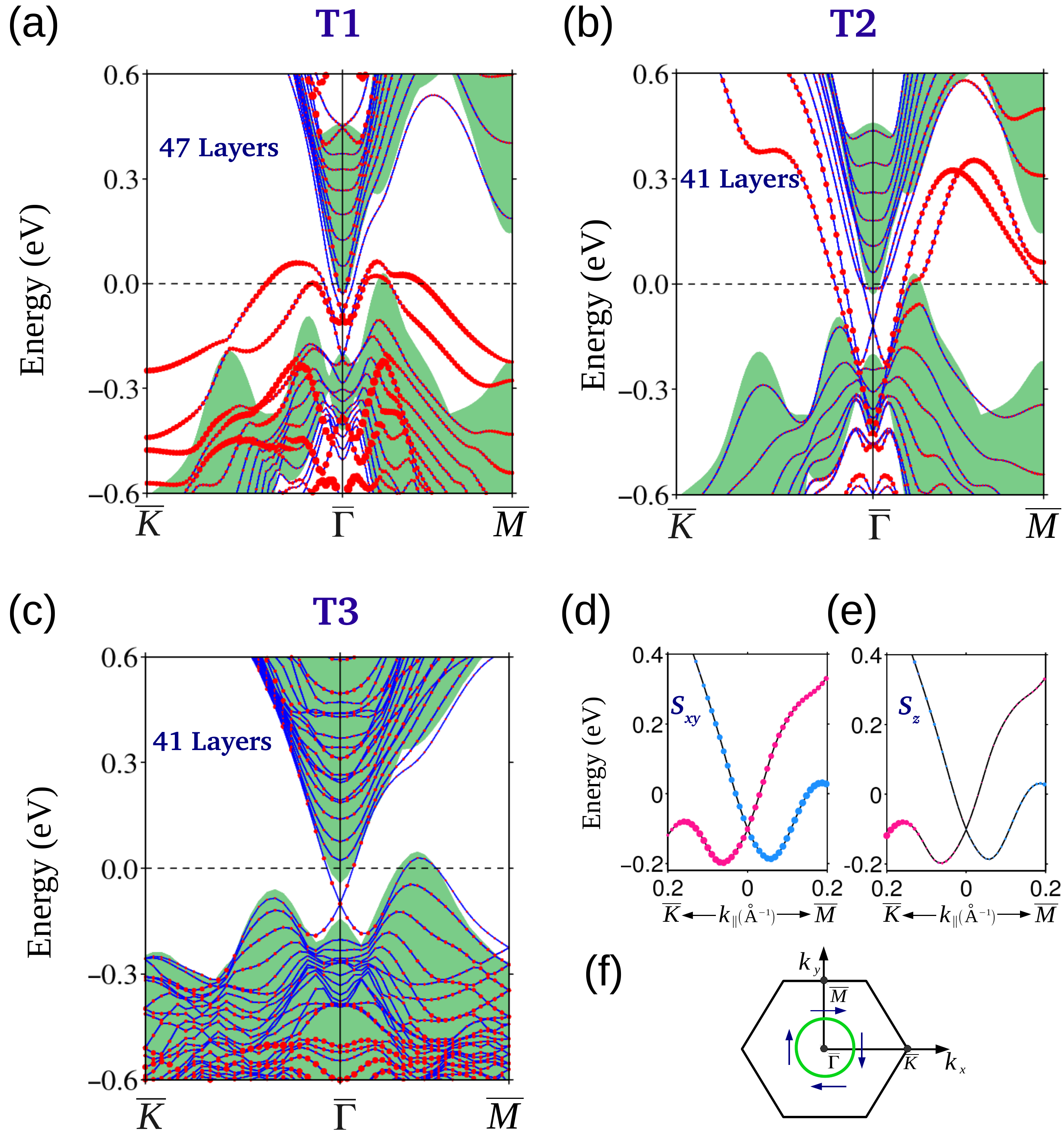} 
\caption{Band structures of symmetric slabs of TlBiTe$_2$ for three different (111) surface terminations: (a) T1 termination with a Te-exposed surface; (b) T2 termination with a Tl-exposed surface; and, (c) T3 termination in which both the Tl- and Te- surfaces are exposed. Color coding is same as in Fig. \ref{fig:tlbise2}. Panels (d) and (e) give the in-plane ($S_{xy}$) and out-of-the-plane ($S_z$) spin-polarizations for the T3 termination. (f) Schematic of the in-plane spin-texture (left-handed) for the surface state electrons.}
\label{fig:tlbite2}
\end{figure}

Band structures of symmetric slabs of TlBiTe$_2$ for T1-T3 terminations are shown in Figs. \ref{fig:tlbite2}(a)-\ref{fig:tlbite2}(c), along with the spin textures of the Dirac cone states for the T3 termination in Figs. \ref{fig:tlbite2}(d)-\ref{fig:tlbite2}(e). We comment only briefly on these results because they are quite similar to those for TlBiSe$_2$ already discussed above, some differences in details notwithstanding. Here also, unlike the T1 and T2 terminations, the T3 termination realizes a single Dirac cone residing within the bulk energy gap without any overlapping trivial surface states, see Fig. \ref{fig:tlbite2}(c), with opposite spin-helicities in the upper and lower portions of the Dirac cone, see Figs.\ref{fig:tlbite2}(d)-\ref{fig:tlbite2}(f). As one moves away from the Dirac node beyond \textbf{k} $\sim$ 0.1 $\r{A}^{-1}$, a significant out-of-plane spin polarization appears and the Dirac cones become hexagonally warped.

On the experimental side, the (111) surface of TlBiTe$_2$ has been investigated via ARPES measurements by Chen et. al.,\cite{TlBiTe2_expchen} who observe a single Dirac cone centered at the $\Gamma$-point without any other overlapping trivial states. The Dirac node lies 0.3 eV below the Fermi level, surrounded by bulk band maxima at \textbf{k} $\sim$ 0.15 $\r{A}^{-1}$ and \textbf{k} $\sim$ 0.2 $\r{A}^{-1}$ along $\overline{\Gamma}-\overline{K}$ and $\overline{\Gamma}-\overline{M}$ directions, respectively. These experimental results are well reproduced by our first-principles computations for the T3 termination. The computed Dirac cone is well-isolated from the trivial surface states, with the Dirac node lying 0.1 eV below the bulk bands and, as shown in Fig. \ref{fig:tlbite2}(c), the Dirac cone is surrounded by projected bulk bands that form maxima at  \textbf{k} $\sim$ 0.15 (0.19) $\r{A}^{-1}$ along $\overline{\Gamma}-\overline{K}$ ($\overline{\Gamma}-\overline{M}$) directions.Notably, here also the total energy per atom of the slab with T3 termination is lower than the average energy per atom of the slabs with T1 and T2 terminations as seen from Table \ref{tab:Energy}, indicating that, like TlBiSe$_2$, T3 provides a good model of the naturally occurring surface of TlBiTe$_2$.

\subsection{Electronic structure of the (221) and (112) surfaces}

\begin{figure}[ht!]
\centering
\includegraphics[width=0.48\textwidth]{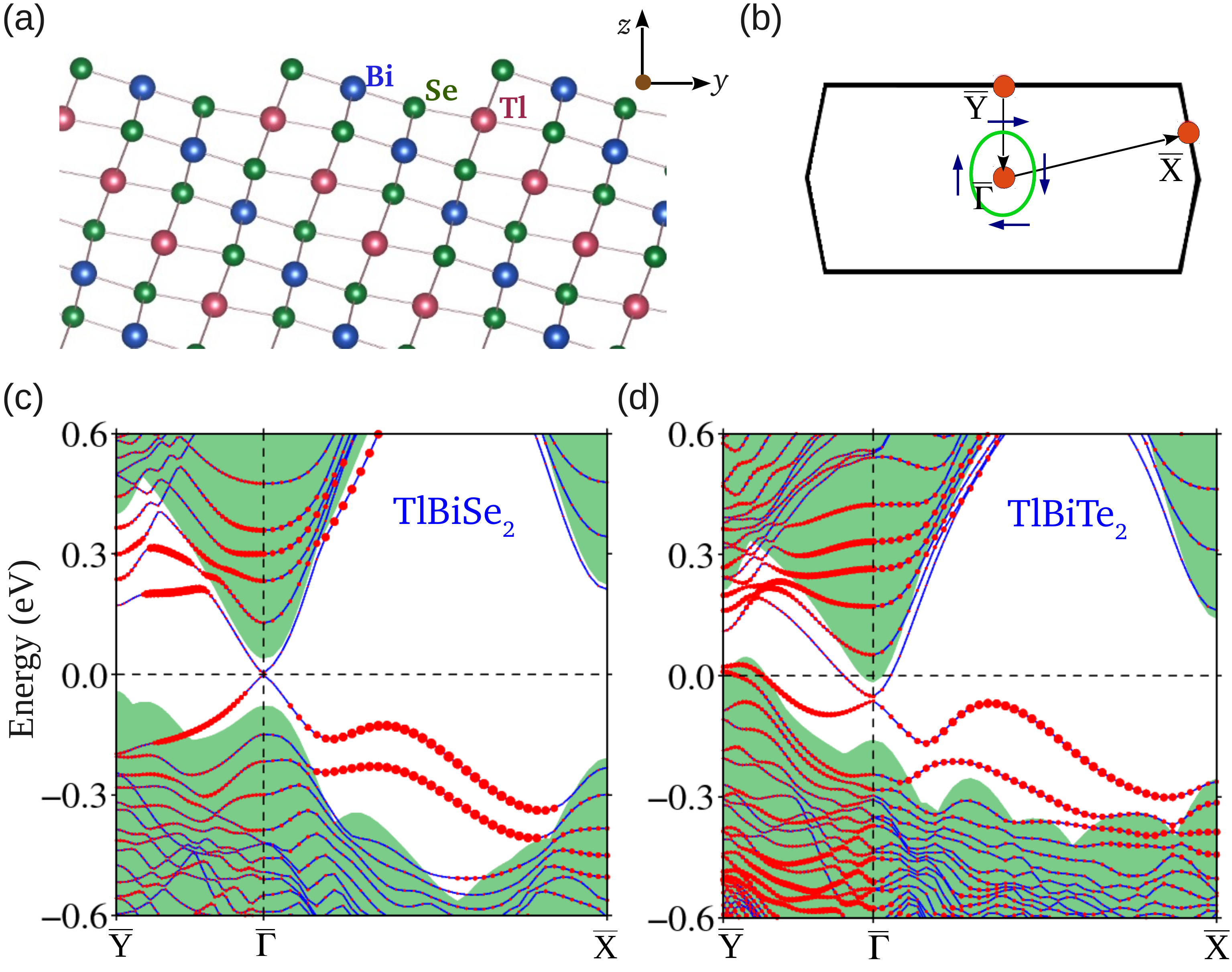} 
\caption{(a) Side view of the (221) surface of TlBiSe$_2$. Se atoms remain at the top layer followed by Bi and Tl layers. (b) Surface Brillouin zone of the (221) surface in which the high symmetry points $\overline{X}$, $\overline{\Gamma}$, and $\overline{Y}$ are marked. Schematic of a constant energy contour with in-plane spin-polarization for the upper Dirac-cone electrons on the (221) surface shown. Band structures of symmetric slabs composed of 111 atomic layers with (c) Se-terminated (221) surface of TlBiSe$_2$ and (d) Te-terminated (221) surface of TlBiTe$_2$. Sizes of red circles are proportional to the weight of states in the top four atomic layers in the slab. The green colored region marks projected bulk bands.}
\label{fig:221surface}
\end{figure}

We have seen that flat surfaces of TlBiSe$_2$ and TlBiTe$_2$ with T1 or T2 type termination produce unwanted (trivial) dangling bond states coexisting with the Dirac cone states, and that this problem can be resolved by invoking the T3 model, which involves a non-flat surface composed of an equal mixture of T1 and T2 terminations. We now discuss another potential route for removing the trivial surface states from the Fermi energy, where the surface remains flat, namely, the (221) surface.Figure \ref{fig:221surface}(a) shows a side view of the (221) surface of TlBiSe$_2$. Normal to the (221) surface makes an angle of $50.6^{\circ}$ with the (111) direction of rhombohedral TlBiSe$_2$, and forms a layered structure like the (111) surface. Unlike the (111) surface, however, the in-plane lattice constants for the (221) surface are unequal ({\it a} = 4.32 \r{A} and {\it b} = 10.06 \r{A}), reducing the symmetry to a single \textit{yz} mirror plane, instead of the three-fold rotation plus a mirror plane symmetry of the (111) surface.\cite{221_theory,221_exp} The (221) surface thus has a strong anisotropy between the \textit{x} and \textit{y} directions, which are symmetric for the (111) surface.

The band structure of a symmetric TlBiSe$_2$ slab composed of 111 atomic layers with a flat (221) surface is shown in Fig. \ref{fig:221surface}(c). Similar to the (111) surface, a single topological surface state with linear energy dispersion is clearly visible within the bulk energy gap. But, unlike the (111) surface with T1 or T2 termination, the trivial surface states are now located below the Dirac node [Fig. \ref{fig:221surface}(c)], leaving the Dirac cone well separated  from the bulk as well as the trivial surface states. Similar results are found with respect to the behavior of the Dirac cone states in the band structure of a TlBiTe$_2$ (221)-terminated slab, shown in Fig. \ref{fig:221surface}(d), although the lower Dirac cone is more distorted and less well isolated from the trivial states in this case compared to the (221)-TlBiSe$_2$ slab. The spin-texture associated with the Dirac cone states on the (221) surfaces of TlBiSe$_2$ and TlBiTe$_2$ are similar to those for the (111) surfaces discussed above, and are not shown in the interest of brevity; the spin-texture for the upper Dirac cone states for the (221)-TlBiSe$_2$ and TlBiTe$_2$ slabs is, however, shown schematically in Fig. \ref{fig:221surface}(b).

\begin{figure}[ht!]
\centering
\includegraphics[width=0.48\textwidth]{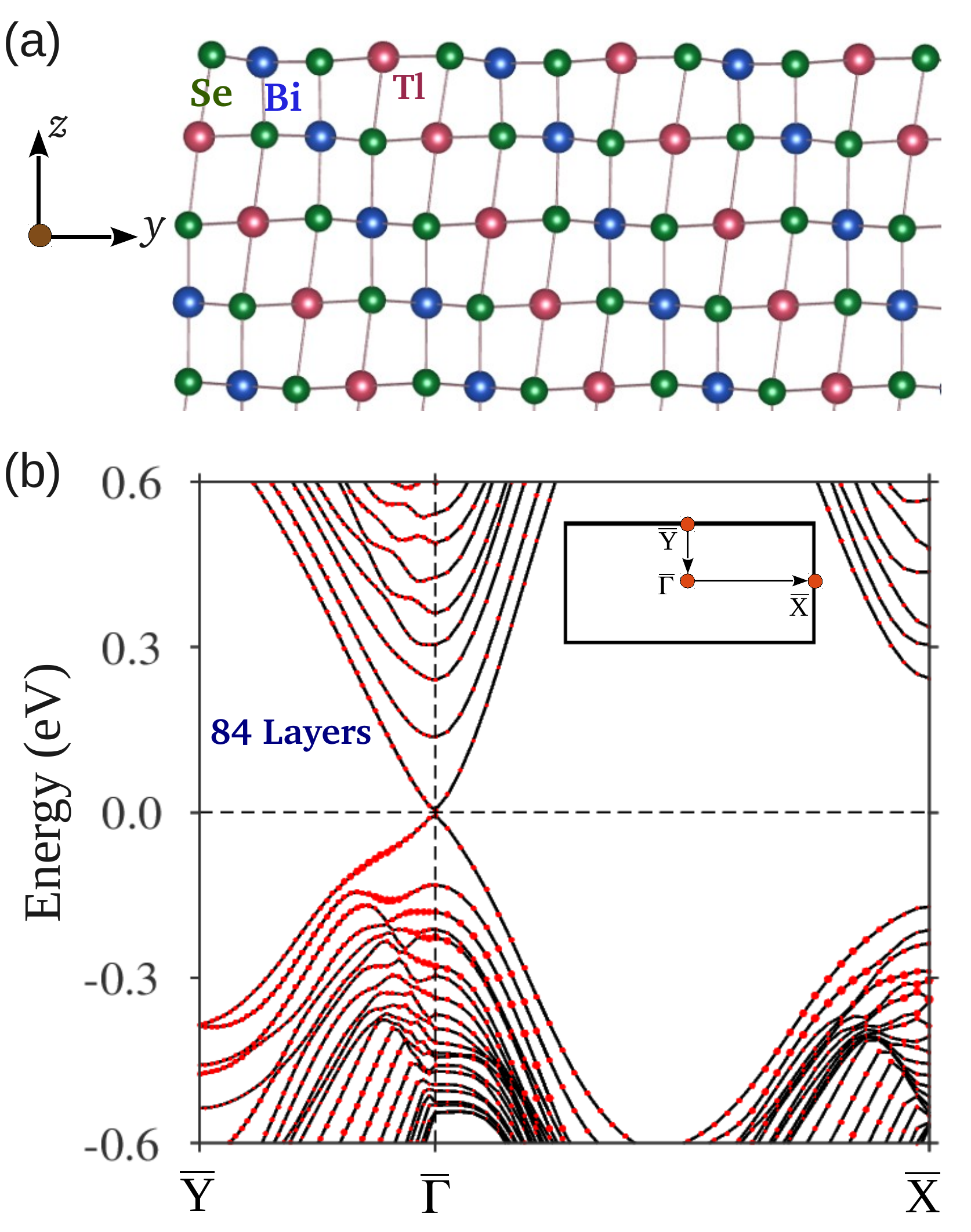} 
\caption{(a) Side view of the atomic structure of the (112) surface of TlBiSe$_2$. (b) Band structure of Se-terminated inversion-symmetric slab of TlBiSe$_2$ (112) surface. Sizes of red circles are proportional to the weight of states in the top four atomic layers in the slab.}
\label{fig:112surface}
\end{figure}

The existence of Dirac cone states over the (221) surface of TlBiSe$_2$, which are well isolated from the bulk as well as the trivial dangling bond states, suggests that other high-index surfaces might also support such a behavior. Accordingly, we have investigated another high-index surface, namely, the (112) surface. The (112) surface has an atomic structure similar to that of the (221) surface, although in this case the normal distance between the top four atomic layers is quite small [see Fig. \ref{fig:112surface}(a)]. As a result, the anion and cation layers lie close to the surface, and we might expect that the strong bonding between these layers will remove the trivial dangling bonds. This is indeed what happens as seen clearly in Fig. \ref{fig:112surface}(b), where an isolated Dirac cone state is found once again within the bulk energy gap with the Dirac node lying at the Fermi level. The trivial surface states have now completely disappeared along the $\overline{\Gamma}-\overline{X}$ directions, which were otherwise located within the bulk energy gap [see Fig. \ref{fig:221surface}(c)].
 
Dirac cone states over high-index surfaces such as (221) and (112) support elliptical constant energy contours, see Fig. \ref{fig:221surface}(a), with anisotropic energy dispersions and velocities of Dirac carriers. This anisotropy is a direct consequence of the reduced symmetry of these surfaces, and results in novel direction dependent properties of the carriers, which would potentially allow manipulation and engineering of new materials platforms for magneto-electronic devices such as anisotropic-magnetic-sensors,\cite{AMR_Nature_park,GMR_yeo,AMR_NL} and use in table-top experiments to simulate high energy particles propagating in anisotropic space.\cite{AMR_Nature_park,GMR_yeo} Other applications based on differences in carrier velocities at the interfaces of surfaces have been suggested.\cite{221_theory}

\section{Conclusion}\label{conclusion}
We have investigated the electronic structures associated with various surface terminations of TlBiSe$_2$ and TlBiTe$_2$ slabs within the framework of the first-principles density functional theory. The Dirac cone states are found to coexist with the trivial dangling bond states for flat (polar) Se/Te- or Tl- terminated surfaces in sharp disagreement with ARPES experiments on TlBiSe$_2$ and TlBiTe$_2$ where no trivial states are observed near the Fermi energy. We show that for a rough (non-polar) (111) surface with an equal number of Tl and Se (Te) atoms in the surface layer of TlBiSe(Te)$_2$, the trivial dangling bond states are removed, leaving a well-isolated Dirac cone in the bulk band gap in remarkable accord with the experimental results. The computed spin-texture of the Dirac states is helical with large in-plane polarization, which is also in agreement with the results of available spin-resolved ARPES experiments. Our study suggests that a rough, nonpolar surface model, such as the T3-termination model, is a viable model of the naturally occurring surfaces of TlBiSe$_2$ and TlBiTe$_2$. Finally, we consider TlBiSe$_2$ and TlBiTe$_2$ slabs with flat, (221) and (112) surfaces, and find that the trivial surface states in this case also tend to move away from the energy region of the bulk band gap, suggesting an alternate route for obtaining Dirac cones which are well-isolated from the bulk and other trivial surface states. Notably, in contrast to the nearly isotropic Dirac states supported by the (111) surfaces of TlBiSe$_2$ and TlBiTe$_2$, the (221) surface Dirac states are more anisotropic, presenting interesting possibilities for fundamental physics as well as applications.

\section*{ACKNOWLEDGMENTS} 
This work was supported by the Department of Science and Technology, New Delhi (India) through project SR/S2/CMP-0098/2010. The work at Northeastern University was supported by the US Department of Energy, Office of Science, Basic Energy Sciences Contract No. DE-FG02-07ER46352, and benefited from Northeastern University's Advanced Scientific Computation Center (ASCC), and the allocation of time at the NERSC supercomputing center through DOE Grant No. DE-AC02-05CH11231. H.L. acknowledges the Singapore National Research Foundation (NRF) for support under NRF Award No. NRF-NRFF2013-03.


\begin{thebibliography}{10}

\bibitem{Moore10}
J. E. Moore, Nature (London) {\bf 464}, 194 (2010).

\bibitem{z2_fu3DTI}
L. Fu, C.~L. Kane, and E.~J. Mele, Phys. Rev. Lett. {\bf 98},  106803  (2007).

\bibitem{z2_kaneIS}
L. Fu and C.~L. Kane, Phys. Rev. B {\bf 76},  045302  (2007).

\bibitem{mzh_RMP1}
M.~Z. Hasan and C.~L. Kane, Rev. Mod. Phys. {\bf 82},  3045  (2010).

\bibitem{review2} X.-L. Qi, S.-C. Zhang, Rev. Mod. Phys. {\bf 83}, 1057 (2011).

\bibitem{Ando13}
Y. Ando, J. Phys. Soc. Jpn. {\bf 82}, 102001 (2013).

\bibitem{bi2se3theory_zhang}
H. Zhang, C.-X. Liu, X.-L. Qi, X. Dai, Z. Fang, and S.-C. Zhang, Nat. Phys. {\bf 5},  438  (2009).

\bibitem{bi2te3_expchen}
Y.~L. Chen, J.~G. Analytis, J.-H. Chu, Z.~K. Liu, S.-K. Mo, X.~L. Qi, H.~J. Zhang, D.~H. Lu, X. Dai, Z. Fang, S.~C. Zhang, I.~R. Fisher, Z. Hussain, and Z.-X. Shen, Science {\bf 325}, 
178  (2009).

\bibitem{bi2se3_expxia}
Y. Xia, D. Qian, D. Hsieh, L. Wray, A. Pal, H. Lin, A. Bansil, D. Grauer, Y.~S. Hor, R.~J. Cava, and M.~Z. Hasan, Nat. Phys. {\bf 5},  398  (2009).

\bibitem{sb2te3_hsieh}
D. Hsieh, Y. Xia, D. Qian, L. Wray, F. Meier, J.~H. Dil, J. Osterwalder, L. Patthey, A.~V. Fedorov, H. Lin, A. Bansil, D. Grauer, Y.~S. Hor, R.~J. Cava, and M.~Z. Hasan, Phys. Rev.
Lett. {\bf 103},  146401  (2009).

\bibitem{TlBiSe2_theoryyan}
B. Yan, C.-X. Liu, H.-J. Zhang, C.-Y. Yam, X.-L. Qi, T. Frauenheim, and S.-C. Zhang, Europhys. Lett. {\bf 90},  37002  (2010).

\bibitem{TlBiSe2_theoryHsin}
H. Lin, R.~S. Markiewicz, L.~A. Wray, L. Fu, M.~Z. Hasan, and A. Bansil, Phys. Rev. Lett. {\bf 105},  036404  (2010).

\bibitem{TlBiSe2_theoryemereev}
S.~V. Eremeev, G. Bihlmayer, M. Vergniory, Y.~M. Koroteev, T.~V. Menshchikova, J. Henk, A. Ernst, and E.~V. Chulkov, Phys. Rev. B {\bf 83},  205129  (2011).

\bibitem{TlBiSe2_theorysingh}
B. Singh, A. Sharma, H. Lin, M.~Z. Hasan, R. Prasad, and A. Bansil, Phys. Rev. B {\bf 86},  115208  (2012).

\bibitem{TlBiSe2_expkuroda}
K. Kuroda, M. Ye, A. Kimura, S.~V. Eremeev, E.~E. Krasovskii, E.~V. Chulkov, Y. Ueda, K. Miyamoto, T. Okuda, K. Shimada, H. Namatame, and M. Taniguchi, Phys. Rev. Lett. {\bf 105}, 
146801  (2010).

\bibitem{TlBiTe2_expchen}
Y.~L. Chen, Z.~K. Liu, J.~G. Analytis, J.-H. Chu, H.~J. Zhang, B.~H. Yan, S.-K. Mo, R.~G. Moore, D.~H. Lu, I.~R. Fisher, S.~C. Zhang, Z. Hussain, and Z.-X. Shen, Phys. Rev. Lett.
{\bf 105},  266401  (2010).

\bibitem{GBT124_theorysingh}
B. Singh, H. Lin, R. Prasad, and A. Bansil, Phys. Rev. B {\bf 88},  195147 (2013).

\bibitem{tetra_Hsin}
H. Lin, T. Das, L.~A. Wray, S.-Y. Xu, M.~Z. Hasan, and A. Bansil, New J. Phys. {\bf 13},  095005  (2011).

\bibitem{all_throughput}
K. Yang, W. Setyawan, S. Wang, M. Buongiorno~Nardelli, and S. Curtarolo, Nat. Mater. {\bf 11},  614  (2012).


\bibitem{qshe_theory}
B.~A. Bernevig and S.-C. Zhang, Phys. Rev. Lett. {\bf 96},  106802  (2006).

\bibitem{spintexture_Roshan}
P. Roushan, J. Seo, C.~V. Parker, Y.~S. Hor, D. Hsieh, D. Qian, A. Richardella, M.~Z. Hasan, R.~J. Cava, and A. Yazdani, Nature {\bf 460},  1106  (2009).

\bibitem{spintronics_ABI}
H. Peng, K. Lai, D. Kong, S. Meister, Y. Chen, X.-L. Qi, S.-C. Zhang, Z.-X. Shen, and Y. Cui, Nat Mater {\bf 9},  225  (2010).

\bibitem{silicene2013Tsai} W.-F. Tsai, C.-Y. Huang, T.-R. Chang, H. Lin, H.-T. Jeng, A. Bansil, Nat Commun {\bf 4}, 1500 (2013).

\bibitem{Magnetic_monopoles}
X.-L. Qi, R. Li, J. Zang, and S.-C. Zhang, Science {\bf 323},  1184  (2009).

\bibitem{Proximity_superconductivity}
L. Fu and C.~L. Kane, Phys. Rev. Lett. {\bf 100},  096407  (2008).

\bibitem{Magnetic_dopingSC}
L.~A. Wray, S.-Y. Xu, Y. Xia, Y.~S. Hor, D. Qian, A.~V. Fedorov, H. Lin, A. Bansil, R.~J. Cava, and M.~Z. Hasan, Nat Phys {\bf 6},  855  (2010).

\bibitem{TPT_TlBiSe2_Xu}
S.-Y. Xu, Y. Xia, L.~A. Wray, S. Jia, F. Meier, J.~H. Dil, J. Osterwalder, B. Slomski, A. Bansil, H. Lin, R.~J. Cava, and M.~Z. Hasan, Science {\bf 332},
  560  (2011).
  
\bibitem{TPT_TlBiSe2_Sato}
T. Sato, K. Segawa, K. Kosaka, S. Souma, K. Nakayama, K. Eto, T. Minami, Y. Ando, and T. Takahashi, Nat. Phys. {\bf 7},  840  (2011).

\bibitem{topologicalfieldtheory}
X.-L. Qi, T.~L. Hughes, and S.-C. Zhang, Phys. Rev. B {\bf 78},  195424 (2008).

\bibitem{dangling}
H. Lin, T. Das, Y. Okada, M. C. Boyer, W. D. Wise, M. Tomasik, B. Zhen, E. W. Hudson, W. Zhou, V. Madhavan, C.-Y. Ren, H. Ikuta, and A. Bansil, 
Nano Letters {\bf 13}, 1915 (2013).

\bibitem{TlBiTe2_super}
R.~A. Hein and E.~M. Swiggard, Phys. Rev. Lett. {\bf 24},  53  (1970).

\bibitem{TlBiSe2_PES}
K. Kuroda, M. Ye, E.~F. Schwier, M. Nurmamat, K. Shirai, M. Nakatake, S. Ueda, K. Miyamoto, T. Okuda, H. Namatame, M. Taniguchi, Y. Ueda, and A. Kimura, Phys. Rev. B {\bf 88},  245308 
(2013).

\bibitem{TlBiSe2_PES1}
F. Pielmeier, G. Landolt, B. Slomski, S. Muff, J. Berwanger, A. Eich, A.~A. Khajetoorians, J. Wiebe, Z.~S. Aliev, M.~B. Babanly, R. Wiesendanger, J. Osterwalder, E.~V. Chulkov, F.~J. Giessibl, and J.~H. Dil, New Journal of Physics {\bf 17},  023067  (2015).

\bibitem{Snte_Hsieh}
T.~H. Hsieh, H. Lin, J. Liu, W. Duan, A. Bansil, and L. Fu, Nat Commun {\bf 3}, 982  (2012).

\bibitem{kohan_dft}
P. Hohenberg and W. Kohn, Phys. Rev. {\bf 136},  B864  (1964).

\bibitem{vasp}
G. Kresse and J. Furthm{\"u}ller, Phys. Rev. B {\bf 54},  11169  (1996).

\bibitem{paw}
G. Kresse and D. Joubert, Phys. Rev. B {\bf 59},  1758  (1999).

\bibitem{pbe}
J.~P. Perdew, K. Burke, and M. Ernzerhof, Phys. Rev. Lett. {\bf 77},  3865
  (1996).

\bibitem{bi2se3theory_Model1}
C.-X. Liu, X.-L. Qi, H. Zhang, X. Dai, Z. Fang, and S.-C. Zhang, Phys. Rev. B {\bf 82},  045122  (2010).

\bibitem{tlbis2_Model2}
B. Singh, H. Lin, R. Prasad, and A. Bansil, Journal of Applied Physics {\bf 116}, 033704 (2014).

\bibitem{BiTeI_ModelSplitt}
M.~S. Bahramy, R. Arita, and N. Nagaosa, Phys. Rev. B {\bf 84},  041202 (2011).

\bibitem{Reconst_theory}
J. Ihm, M.~L. Cohen, and D.~J. Chadi, Phys. Rev. B {\bf 21},  4592  (1980).

\bibitem{Reconst_exp}
M. McEllistrem, M. Allgeier, and J.~J. Boland, Science {\bf 279},  545  (1998).

\bibitem{footnote1}
Note that our 2$\times$2 supercell construction is not equivalent to having a c(2$\times$2) supercell. It is easily seen, for example, that the c(2$\times$2) supercell will not break the in-plane bulk lattice symmetry, and yield only T1 and T2 termination.

\bibitem{footnote2}
Ref. \citenum{TlBiSe2_theoryemereev} considers a Tl-Se swap model in which the surface Tl layer is interchanged with the underlying Se layer to support Dirac cone states in agreement with experiments. However, this model was disregarded owing to its higher total energy compared to the original Tl-terminated T2 surface.

\bibitem{221_theory}
C.-Y. Moon, J. Han, H. Lee, and H.~J. Choi, Phys. Rev. B {\bf 84},  195425 (2011).

\bibitem{221_exp}
Z. Xu, X. Guo, M. Yao, H. He, L. Miao, L. Jiao, H. Liu, J. Wang, D. Qian, J. Jia, W. Ho, and M. Xie, Advanced Materials {\bf 25},  1557  (2013).

\bibitem{AMR_Nature_park}
C.-H. Park, L. Yang, Y.-W. Son, M. L. Cohen, and S. G. Louie, Nature Physics {\bf 4}, 213 (2008). 

\bibitem{GMR_yeo}
Z. Yue, X. Wang, and S. Dou, Integrated Ferroelectrics {\bf 140}, 155 (2012).

\bibitem{AMR_NL}
S. Tang and M. S. Dresselhaus, Nano Letters {\bf 12}, 2021 (2012).


\end{thebibliography}
\end{document}